\documentclass[aps,prl,twocolumn,nofootinbib,superscriptaddress,showpacs,tightenlines]{revtex4}
\usepackage{amsfonts}
\usepackage{amsmath}
\usepackage{amssymb}  
\usepackage{hyperref}
\usepackage{stmaryrd}

\def\be{\begin{equation}}
\def\ee{\end{equation}}
\def\ba{\begin{eqnarray}}
\def\ea{\end{eqnarray}}

\def\lp{\ell_\text{Pl}}

\def\tr{\text{tr}}

\def\de{\mathrm{d}}

\def\SU{\text{SU}}

\def\SL{\text{SL}}
\def\su{\mathfrak{su}}

\def\ut#1{\rlap{\lower1ex\hbox{$\sim$}}#1{}}
\newcommand{\N}{\mathbb{N}}
\newcommand{\C}{\mathbb{C}}
\newcommand{\R}{\mathbb{R}}
\DeclareFontFamily{U}{rsfs}{}         
\DeclareFontShape{U}{rsfs}{m}{n}{<5> rsfs5 <6><7> rsfs7          %
  <8><9><10><10.95><12><14.4><17.28><20.74><24.88> rsfs10}{}     %
\DeclareMathAlphabet{\mathfs}{U}{rsfs}{m}{n}                     %
\newcommand{\mfs}[1]{\mathfs {#1}}                               %

\newcommand{\sA}{{\mfs A}}

\begin{document}

\title{Black Hole Entropy from complex Ashtekar variables}

\author{Ernesto Frodden}
\affiliation{Centre de Physique Th\'eorique\footnote{Unit\'e Mixte de Recherche (UMR 6207) du CNRS et des Universit\'es Aix-Marseille I, Aix-Marseille II, et du Sud Toulon-Var; laboratoire afili\'e \`a la FRUMAM (FR 2291).}, Campus de Luminy, 13288 Marseille, France.}

\author{Marc Geiller\footnote{New address: Institute for Gravitation and the Cosmos \& Physics Department, Penn State, University Park, PA 16802, U.S.A.}}
\affiliation{Laboratoire APC -- Astroparticule et Cosmologie, Universit\'e Paris Diderot Paris 7, 75013 Paris, France.}

\author{Karim Noui}
\affiliation{Laboratoire de Math\'ematique et Physique Th\'eorique\footnote{F\'ed\'eration Denis Poisson Orl\'eans-Tours, CNRS/UMR 6083.}, 37200 Tours, France.}
\affiliation{Laboratoire APC -- Astroparticule et Cosmologie, Universit\'e Paris Diderot Paris 7, 75013 Paris, France.}

\author{Alejandro Perez}
\affiliation{Centre de Physique Th\'eorique\footnote{Unit\'e Mixte de Recherche (UMR 6207) du CNRS et des Universit\'es Aix-Marseille I, Aix-Marseille II, et du Sud Toulon-Var; laboratoire afili\'e \`a la FRUMAM (FR 2291).}, Campus de Luminy, 13288 Marseille, France.}

\begin{abstract}
In loop quantum gravity, the number $N_\Gamma(A,\gamma)$ of microstates of a black hole for a given discrete geometry $\Gamma$ depends on the so-called Barbero-Immirzi parameter $\gamma$. Using a suitable analytic continuation of $\gamma$ to complex values, we show that the number $N_\Gamma(A,\pm i)$ of microstates behaves as $\exp(A/(4\lp^2))$ for large area $A$ in the large spin semiclassical limit. Such a correspondence with the semiclassical Bekenstein-Hawking entropy law points towards an unanticipated and remarkable feature of the original complex Ashtekar variables for quantum gravity.
\end{abstract}


\maketitle

The Barbero-Immirzi parameter $\gamma$ was originally introduced \cite{barbero} as a way to circumvent the problem of imposing the reality constraints in the complex (self-dual) Ashtekar formulation of gravity \cite{ashtekar}. Historically, $\gamma$ appeared as a parameter labeling a family of canonical transformations turning the ADM phase space into the so-called Ashtekar-Barbero phase space, parametrized by a real $\su(2)$ connection and its conjugate momentum. Later on, Holst \cite{holst} realized that this Hamiltonian formulation of gravity could be obtained by adding to the standard Hilbert-Palatini Lagrangian a topological term with $\gamma$ as a coupling constant. This term vanishes due to the Bianchi identities when one resolves the spin connection in terms of the tetrad, and for this reason $\gamma$ is not relevant at the classical level.

In the quantum theory however, the Barbero-Immirzi parameter plays a crucial role since the spectrum of the geometric operators is discrete in units of the loop quantum gravity (LQG) scale $\ell_\text{LQG}=\sqrt{\gamma G\hbar}=\sqrt{\gamma}\lp$, where $\lp$ is the Planck length \cite{rovelli-smolin}. Moreover, this $\gamma$-dependency of the fundamental physical cut-off is inherited by the value of the black hole entropy in the LQG calculation. Compatibility with the expected semiclassical value $S=A/(4\lp^2)$ (where $A$ is the area of the horizon) requires that $\gamma$ be fixed to a particular real value. In fact, a lot of different techniques have been developed in order to obtain the value of $\gamma$ \cite{BHentropy1,We1} (see also \cite{livine-terno}).

In LQG, the horizon of a black hole has the topology of a 2-sphere, with colored punctures coming from the spin network links that cross the horizon. Each puncture carries a quantum of area, and the sum of these microscopic areas gives the macroscopic area $A$ of the horizon. In the microcanonical ensemble, the entropy of the black hole is given by the logarithm of the number $N(A,\gamma)$ of microscopic states compatible with  the macroscopic area $A$ of the horizon. As suggested by the notation, the quantity $N(A,\gamma)$ depends on $A$ but also on $\gamma$, since admissible microstates are sets $\Gamma=\{j_1,\cdots, j_p\}$ of punctures labelled by spins, where $p$ is the number of punctures, satisfying the quantum area constraint
\be
A-\epsilon<8\pi\gamma\lp^2\sum_{\ell=1}^p\sqrt{C(j_\ell)}<A+\epsilon,
\ee
for some small coarse graining $\epsilon>0$, and $C(j)={j(j+1)}$.

It has been shown recently \cite{We1} that there is a close relationship between black holes in LQG and $\SU(2)$ Chern-Simons theory. In this framework, the number $N(A,\gamma)$ of microstates can be expressed as
\be\label{sum}
N(A,\gamma)=\sum_\Gamma w_\Gamma N_\Gamma(A,\gamma),
\ee
where $N_\Gamma(A,\gamma)$ is the dimension of the Hilbert space of $\SU(2)$ Chern-Simons theory on a punctured 2-sphere, and $w_\Gamma$ is a weight assigned to each $\Gamma$.


However, the present state of development of LQG is not conclusive about the precise form of the weights $w_\Gamma$, and one must admit that this remains to a large extend an open issue. It seems clear to us that these weights should be fixed by dynamical considerations, in relation in particular to the requirement that a semiclassical geometry plus field configuration (the suitable low energy vacuum state) be recovered near the horizon. In four dimensions, some first steps in this direction have been explored in \cite{Ghosh:2012wq}. In fact, perhaps the most transparent example of the dynamical nature of the weights $w_\Gamma$ is the three-dimensional pure gravity description of the BTZ black hole \cite{btz}. In four dimensions, we will show in this letter that the key result concerning the analytic continuation, together with the assumption that punctures are indistinguishable \cite{kay}, lead to a complete semiclassical agreement with Bekenstein-Hawking thermodynamics.


More precisely, we are going to propose an expression for $N_\Gamma(A,\pm i)$, which is the number of states for a theory defined in terms of the complex self-dual Ashtekar connection, i.e. for $\gamma=\pm i$. Our proposal is based on a certain analytic continuation of the formula for the dimension of the Hilbert space of $\SU(2)$ Chern-Simons theory from $\SU(2)$ representations to suitable $\SL(2,\C)$ representations satisfying self-duality constraints. The striking result is that after analytic continuation the asymptotic behavior of the Chern-Simons Hilbert space is holographic, i.e. satisfies
\be
\log\big(N_\Gamma(A,\pm i)\big)\stackrel{\text{s.c.}}\sim\frac{A}{4\lp^2},
\ee
where $\stackrel{\text{s.c.}}\sim$ denotes that the result is valid in the large spin ({\em semiclassical}) asymptotic regime. This result suggests that the complex formulation in terms of the Ashtekar variables (which is to be put in parallel with the so-called Lorentz-covariant formulation \cite{alexandrov1}) could lead to a clear-cut derivation of the black hole entropy in the framework of quantum gravity.

\subsubsection*{Analytic continuation}

Black holes in LQG can be described in terms of an $\SU(2)$ Chern-Simons theory
\be\label{CSaction}
S_\text{CS}[\sA]=\frac{k}{4\pi}\int_\Delta\tr\left(\sA\wedge\de \sA+\frac{2}{3}\sA\wedge \sA\wedge \sA\right),
\ee
where $k\propto A$ (the precise form of the level will not be important), and $\Delta$ is the black hole horizon with spatial area $A$. This area $A$ is obtained as the sum of the fundamental contributions carried by the $p$ links $\ell$ crossing the horizon, according to the formula
\be\label{area rel}
A=8\pi\lp^2\gamma\sum_{\ell=1}^p\sqrt{C(j_\ell)},
\ee
where $\sqrt{C(j_\ell)}$ is the quantum of area associated to the puncture $\ell\in\llbracket1,p\rrbracket$, which is colored with an $\SU(2)$ representation of spin $j_\ell$ and of finite dimension $d_\ell=2j_\ell+1$. Usually, $C(j)=j(j+1)$ is the quadratic Casimir operator evaluated in the representation $j$, but there exists also a linear model with $C(j)=j^2$, which corresponds to a different regularization of the area operator. These two expressions obviously agree for large spins.

The Hilbert space associated to a black hole whose horizon is crossed by $p$ colored links, is exactly the Hilbert space of the $\SU(2) $ Chern-Simons theory (\ref{CSaction}) with $p$ punctures colored with the representations of spins $(j_1,\dots,j_p)$. Its dimension is therefore given by the finite sum \cite{KM,ENPP}
\ba\label{13}
&&\phantom{=}D_k(j_1,\dots,j_p)\label{dim}\nonumber\\
&&=\frac{2}{k+2}\sum_{d=1}^{k+1}\sin^2\left(\frac{\pi d}{k+2}\right)\prod_{\ell=1}^p\frac{\displaystyle\sin\left(\frac{\pi dd_\ell}{k+2}\right)}{\displaystyle\sin\left(\frac{\pi d}{k+2}\right)}.
\ea

A configuration for a black hole of horizon area $A$ is an assignment of $p$ spins $(j_1,\dots,j_p)$ compatible with $A$ in the sense that the relation (\ref{area rel}) is satisfied. As a consequence of this restriction, the number of microstates for a given puncture configuration is a function of the Barbero-Immirzi parameter as well. To remind us of this fact we introduce the notation 
\be\label{states}
N_\Gamma(A,\gamma)={D_k(j_1,\dots,j_p)}.
\ee 
The total number of microstates is given by a sum of the type (\ref{sum}) over the colored graphs $\Gamma$.

The analytic continuation that we propose consists in replacing the spins $j$ by complex values according to the rule
\be
j_\ell\to is_\ell-\frac{1}{2},
\ee
where $s\in \R$. As we are going to show below, this corresponds in a precise sense to choosing self-dual representations, i.e., solutions of the self-duality constraints which in addition satisfy suitable {\em reality conditions}. Using the analyticity of $D_k(j_1,\dots,j_p)$ in its arguments, the above transformation defines uniquely the number density $N^{\text{\tiny A}}_\Gamma$ that we are looking for. Namely,
\ba\label{dimcont}
&&\phantom{=}D_k(j_1,\dots,j_p) \to  i^{p} D_k(is_1-\frac{1}{2},\dots,is_p-\frac{1}{2})\nonumber\\
&&=\frac{2}{k+2}\sum_{d=1}^{k+1}\sin^2\left(\frac{\pi d}{k+2}\right)\prod_{\ell=1}^p\frac{\displaystyle\sinh\left(\frac{2 \pi d s_{\ell}}{k+2}\right)}{\displaystyle\sin\left(\frac{\pi d}{k+2}\right)},
\ea
where the factor $i^{p}$ comes from the Jacobian $\rm{d}\mathbf{j}/\rm{d}\mathbf{s}$ necessary when dealing with densities. When the spins $s_\ell$ are large, this sum is dominated by the exponential with the largest argument (obtained for $d=k+1$), which is given by
\ba\label{dim i}
 &&\phantom{\approx}N_\Gamma(A,\pm i)=i^{p}D_k(is_1-\frac{1}{2},\dots,is_p-\frac{1}{2})\nonumber\\
 &&\approx\frac{1}{k}\sin\left(\frac{\pi(k+1)}{k+2}\right)^{2-p}\prod_{\ell=1}^p\sinh(2\pi s_\ell).
\ea
This is the main equation of the paper\footnote{In a previous version of this work, a different version of the analytic continuation has been proposed, where 
instead of the present procedure the replacement $k\to ik$ in performed (\ref{13}). In order for this to make mathematical sense, one would need to justify several issues. First, the upper bound in the sum over the dummy variable $d$ has to remain real if the dummy variable is to remain real. For this one should argue that the upper bound is actually given by $|k|+1$ instead of $k+1$. Second, in order not to get an obviously complex result for the number of states, one would also need to replace the overall factor of $1/(k+2)$ by $1/(|k|+2)$. With this, the result would still be complex because of the $k+2$ dependence of the $\sin$ functions in (\ref{13}). Only the leading order would be real and positive in the large $|k|$ asymptotic expansion. With all these ingredients, the result was
\ba
&&\phantom{=}D_k(j_1,\dots,j_p)\to D_{ik} (j_1,\dots,j_p)\nonumber\\
&&=\frac{2}{|k|+2}\sum_{d=1}^{|k|+1}\sin^2\left(\frac{\pi d}{ik+2}\right)\prod_{\ell=1}^p\frac{\displaystyle\sin\left(\frac{2 \pi d d_{\ell}}{i k+2}\right)}{\displaystyle\sin\left(\frac{\pi d}{i k+2}\right)}\nonumber\\
&&\approx\frac{2}{|k|}\sum_{d=1}^{|k|}\sinh^2\left(\frac{\pi d}{k}\right)\prod_{\ell=1}^p\frac{\displaystyle\sinh\left(\frac{2 \pi d d_{\ell}}{k}\right)}{\displaystyle\sinh\left(\frac{\pi d}{ k}\right)}.
\ea
Interestingly, this formula has appeared in the description of the entropy of a BTZ black hole from the canonical point of view \cite{btz} and from the point of view of spin foam models \cite{btz-tv}, where in order to get a negative cosmological constant it is indeed the level of Chern-Simons theory which has to be analytically-continued.}.

It is important to point out that our procedure of analytic continuation (\ref{dimcont}) would be a rigorous mathematical fact if the starting point was a function of $p$ real variables. In this case the dimension would extended uniquely to a function of $p$ complex variables. However, as the input is a function of $p$ integer variables $d_\ell$, the analytic continuation is ambiguous. For instance, a different result would be obtained by multiplying the input by any function equal to one when the entries are integers. What makes our result unique is that only the expression (\ref{13}) with complex variables as entries can be given a field theoretical description in the context of the analytic continuation of Chern-Simons theory \cite{Witten}.

\subsubsection*{Self dual representations and reality conditions}
\label{SFS}

Now, we clarify the meaning of the representations  $s_\ell$ appearing in formula \eqref{dim i}. In the first place, they are self-dual or anti self-dual representations of $\SL(2,\mathbb{C})$ solutions to the constraints  
\be\label{reality}
\vec{L}\pm i\vec{K}=0,
\ee
where $\vec{L}$ and $\vec{K}$ are the generators of rotations and boost in the internal $\SL(2,\C)$ group. For simplicity, let us focus from now on on the self-dual representations, i.e., with a plus sign in the previous constraint. From the Lie algebra $\mathfrak{sl}(2,\C)$, one can immediately show that the three equations in (\ref{reality}) are first class constraints. Thus, it makes sense to look for solutions of these constraints among representations of $\SL(2,\C)$. If we label, as usual, by $(p,k)$ the irreducible representations of the Lorentz group, then the previous condition (\ref{reality}) is solved strongly in different ways.

One has on the one hand solutions which are finite dimensional and simply correspond to the usual spinor representations of the Lorentz group. These solutions are found by expanding $\SL(2,\C)$ representations in terms of unitary irreducible representations of the subgroup $\SU(2)$. They are the ones used in standard field theory to describe matter fields. However, they must be discarded in the present gravitational context because they fail to satisfy some necessary reality requirements imposed by Ashtekar's self-dual formulation. More precisely, a simple analysis shows that the spectrum of the square of the flux operators is negative-definite (the area spectrum is imaginary). This is because
\ba\label{12}
\widehat{\text{Area}}^2|j\rangle&=&\widehat{\Sigma}_{\pm}\cdot\widehat{\Sigma}_{\pm}|j \rangle 
=-(8\pi\lp^2)^2j(j+1)|j\rangle,
\ea
where $|j\rangle$ denotes a flux excitation, and $\Sigma_\pm$ is the (anti) self-dual component of the densitized triad field ($\Sigma=e\wedge e$). The appearance of the uncomfortable minus sign in (\ref{12}) is a well known fact since the very early stages of construction of the LQG formalism \cite{Rovelli:1994ge}, and one of the reasons for using the real connection variables with a real Barbero-Immirzi parameter.
  
However, there are other solutions of (\ref{reality}) which become apparent when expanding $\SL(2,\C)$ representations in terms of unitary irreducible representations of the subgroup $\SU(1,1)$ (as in \cite{Conrady:2010sx}). The group $\SU(1,1)$ has two types of unitary representations: those that are discrete and labelled by $j\in \N/2$, and those in the so-called principal series and labelled by $j=is-1/2$. There are solutions of the self-duality constraints (\ref{reality}) in one-to-one correspondence with both types of solutions. The ones associated with the discrete series fail (just like the finite-dimensional ones) the reality condition (\ref{12}). However, those associated with the principal series lead to a {\em real} (yet continuous) area spectrum:
\be\label{area}
\widehat{\text{Area}}|s\rangle=
8\pi\lp^2\sqrt{s^2+1/4}|s\rangle\approx8\pi\lp^2s|s\rangle,
\ee
where in the last line we have written a large $s$ approximation of the eigenvalues.

The members of the principal series are the solutions we are looking for. From the point of view of the representations of $\SL(2,\C)$, they correspond to labels $\chi=(p,k)$ such that $p=-2i(j+1)$ and $k=\mp 2ij$, or $p=2ij$ and $k=\pm 2(j+1)$ \cite{SFi}, where $j=is-1/2$ and $s\in\R$. All this justifies the analytic continuation of the previous section and provides a clear interpretation of equation (\ref{dim i}). Assuming that $k$ is a non dynamical regulator, then in the large $s_{\ell}$ limit we have
\be\label{hb}
\log\big(N_\Gamma(A,\pm i)\big)\approx {\frac{A}{4\lp^2}},
\ee
where $A$ in $N_\Gamma(A,\pm i)$ is the area eigenvalue of the state with $p$ punctures. The special role of $\SU(1,1)$ representations in the principal series in three dimensions is clarified \cite{BGNY}.

The above formula exhibits a {\em holographic} property of the number of states $N_\Gamma(A,\pm i)$, at least in the large $s_{\ell}$ limit. Remarkably, this is enough to recover the Bekenstein-Hawking entropy formula when summing over states $\Gamma$, if one adds the additional (but very natural) assumption that punctures are indistinguishable excitations of the gravitational field \cite{kay}. It follows from simple statistical mechanical considerations that such system can only be macroscopic if its temperature is $T=T_\text{Hw}\big(1+o(\lp/\sqrt{A})\big)$ (with $T_\text{Hw}$ the Hawking temperature), and that the number of punctures is $p\approx\sqrt{A}/\lp$, which in turn implies that the system is dominated by large quantum numbers $s_{\ell}$, and finally that the entropy is
\be
S_\text{BH}\equiv\log\big(N_\Gamma(A,\pm i)\big)={\frac{A}{4\lp^2}}\big(1+o({\lp}/\sqrt{A})\big).
\ee
Notice that the entropy is finite despite the fact that the labels $s_{\ell}$ are continuous variables to be integrated over.

\subsubsection*{Discussion}


Our result suggests a possible way towards the quantization of Ashtekar gravity, that would mimic our procedure for the quantum black hole. Such a recipe could consist in first starting from the kinematical quantum theory in terms of the real Ashtekar-Barbero connection (leading to a well understood Hilbert space structure based on the notion of spin network states and quantum geometry) in order,  in a second step, to produce physical quantities by means of an analytic continuation of suitable quantities along the lines presented here. A particularly interesting question is that of the definition of {\em spin foam} transition amplitudes by means of such a procedure \cite{SFi}.

Our results open important and (probably quite difficult) mathematical questions that require a systematic analysis. In particular, the representations selected by the self-duality condition (\ref{reality}) and the reality conditions are exotic representations from the point of view of $\SL(2,\C)$. The representations $\chi_1=(-2i(j+1),\pm 2ij)$ and  $\chi_1=(-2i j,\pm 2i(j+1))$, with $j=is-1/2$ are not only non-unitary but act on the space of homogeneous functions $f(z_1,z_2)$ which are non-analytic (they have branch-cut singularities that allow $k$ to be different from an integer). We believe that a deeper understanding of these representations will clarify further the geometrical meaning of our analytic continuation. This is certainly not an easy task and it represents a longer term objective of our work. Another important question to be studied in the future is the relationship of our results with those of Bianchi \cite{bianchi} and Pranzetti \cite{Pranzetti:2013lma}.

\subsubsection*{Acknowledgements}

We would like to thank Abhay Ashtekar and Muxin Han for discussions.

\end{document}